\documentstyle[12pt]{article}
\title{
Does the Falicov-Kimball model allow for a ferroelectric ground state with
a spontaneous polarization?}
\author{Pavol Farka\v sovsk\'y\\
Institute  of  Experimental  Physics,  Slovak   Academy   of
Sciences\\
Watsonova 47, 043 53 Ko\v {s}ice, Slovakia}
\date{}
\begin{document}
\baselineskip=24pt
\maketitle

\begin{abstract}

The extrapolation of small-cluster exact-diagonalization calculations
is used to examine the possibility of electronic ferroelectricity
in the one dimensional spinless and spin-one-half Falicov-Kimball
model (FKM). It is found that neither spinless nor spin-one-half version
of the FKM does not allow for a ferroelectric ground state with a
spontaneous polarization, i.e. there is no nonvanishing
$<d^{+}f>$-expectation value for vanishing hybridization $V$.

\end{abstract}

\newpage

\section{Introduction}
In the last few years the FKM~\cite{Falicov} has been extensively studied in
connection with the exciting idea of electronic ferroelectricity~\cite{P1,P2,
Cz}. It is generally supposed that the ferroelectricity in mixed-valent
compounds is of purely electronic origin, i.e. it results from an electronic
phase transition, in contrast to the conventional displacive ferroelectricity
due to a lattice distortion. Since the FKM is probably the simplest model of
electronic phase transitions in rare-earth  and transition-metal compounds
it was natural to test the idea of electronic ferroelectricity just
on this model.

The FKM is based on the coexistence of two
different types of electronic states in a given material:
localized, highly correlated ionic-like states and extended,
uncorrelated, Bloch-like states. It is accepted that
insulator-metal transitions result from a change in the occupation
numbers of these electronic states, which remain themselves basically
unchanged in their character. Taking into account only the
intra-atomic Coulomb interaction between the two types of states,
the Hamiltonian of the spinless FKM with hybridization
can be written as the sum of four terms:

\begin{equation}
H=\sum_{ij}t_{ij}d^+_id_j+U_{df}\sum_if^+_if_id^+_id_i+E_f\sum_if^+_if_i+
V\sum_id^+_if_i+h.c.,
\end{equation}
where $f^+_i$, $f_i$ are the creation and annihilation
operators  for an electron in  the localized state at
lattice site $i$ with binding energy $E_f$ and $d^+_i$,
$d_i$ are the creation and annihilation operators
of the itinerant spinless electrons in the $d$-band
Wannier state at site $i$.

The first term of (1) is the kinetic energy corresponding to
quantum-mechanical hopping of the itinerant $d$ electrons
between sites $i$ and $j$. These intersite hopping
transitions are described by the matrix  elements $t_{ij}$,
which are $-t$ if $i$ and $j$ are the nearest neighbors and
zero otherwise (in the following all parameters are measured
in units of $t$). The second term represents the on-site
Coulomb interaction between the $d$-band electrons with density
$n_d=\frac{1}{L}\sum_id^+_id_i$ and the localized
$f$ electrons with density $n_f=\frac{1}{L}\sum_if^+_if_i$,
where $L$ is the number of lattice sites. The third  term stands
for the localized $f$ electrons whose sharp energy level is $E_f$.
The last term represents the hybridization between the itinerant
and localized states.

In spite of fact that many of ground-state properties of the spinless FKM
(the nature of the ground state~\cite{Gr}, the picture of valence and
metal-insulator transitions~\cite{F1,F2}, etc.) are well understood at present,
the problem of electronic ferroelectricity  remains still
an open question. Very recently Portengen at al.~\cite{P1,P2} studied the
FKM with a k-dependent hybridization in Hartree-Fock approximation
and found, in particular, that a non-vanishing excitonic
$<d^{+}f>$-expectation value exists even in the limit of vanishing
hybridization $V \rightarrow 0$. As an applied (optical) electrical field
provides for excitations between d- and f-states and thus for a polarization
expectation value $P_{df}=<d_i^{+}f_i>$, the finding of a spontaneous
$P_{df}$ (without hybridization or electric field) has been interpreted as
evidence for electronic ferroelectricity.
However, analytical calculations within well controlled approximation (for
$U_{df}$ small) performed by Czycholl~\cite{Cz} in infinite dimensions do
not confirm this conclusion. In contrast to results obtained by Portengen et
al.~\cite{P1,P2} he found that the symmetric ($E_f=0, n_f=n_d=0.5$) FKM does
not allow for a  ferroelectric ground state with a spontaneous polarization,
i.e. there is no nonvanishing $<d^{+}f>$-expectation value in the
limit of vanishing hybridization.

In order to shed some light on this controversy we have decide to study the
problem of electronic ferroelectricity in the FKM using the small-cluster
exact-diagonalization method that was so successful in describing
of valence and metal-insulator transitions in this model~\cite{F1,F2}.
It should be noted that for given $L$ the full Hilbert space of the
spinless FKM consists of $4^L$ quantum states,
thereby strongly limiting numerical computations.
Although the number of states can be reduced considerably by the use
of symmetries of $H$, there is still a limit $(L \sim 10)$ on the size of
clusters that can be studied using small-cluster exact-diagonalization
calculations.
However, we will show later that due to small sensitivity of the FKM
on $L$ (for a wide range of parameters), already such small clusters can
describe satisfactory the behavior of the $<d^+f>$-expectation value
that lies in the center of our interest.

To compare our numerical results with Czycholl`s ones~\cite{Cz} obtained 
for small Coulomb interactions we have started our investigation in the 
weak-coupling limit and half-filled band case ($n_f=n_d=0.5$). 
The weak-coupling numerical results for $P_{df}$ obtained using the
modified Lanczos method~\cite{Lan,F3} are displayed in Fig.~1. To reveal the
finite-size  effects on $P_{df}$ numerical calculations have been performed
for three finite clusters of $L=6,8$ and 10 sites. It is seen that there are
nonzero  finite-size effects on the $<d^+f>$-expectation value in the
weak-coupling limit, however they do not change qualitatively the behavior of
$P_{df}$ in the limit $V\to 0$ that is crucial for the verification of
spontaneous polarization. In all cases the $<d^+f>$-expectation value vanishes
in the limit $V\to 0$, so there is no spontaneous polarization in the spinless
FKM. Thus, in accordance with Czycholl`s weak-coupling results we can conclude
that the spinless FKM does not allow for a ferroelectric ground state
with a spontaneous polarization at least for $U_{df}$ small.

Unlike the method used by Czycholl that is restricted to small interactions
we can proceed in the numerical study of the FKM at arbitrary $U_{df}$.
The strong-coupling numerical results for $P_{df}$ are displayed in Fig.~1
for $U_{df}=4$ and $U_{df}=10$.
Obviously the one-dimensional FKM does not exhibit a ferroelectric
ground state with a spontaneous polarization in the strong-coupling
limit. For both values of $U_{df}$  the $<d^+f>$-expectation value
vanishes for $V\to 0$, and it is demonstrated that this result is independent
of $L$. Thus, the strong-coupling results can be satisfactory extended to
large systems and they should be considered as definite.

Of course, the absence of ferroelectric ground state in the spinless
FKM does not exclude that some other models could exhibit such a ground
state. The spinless FKM is not too realistic model of
a rare-earth compound, because any real Fermi system has at least
a spin degeneracy. Therefore, it is natural to ask if the spin-one-half
FKM would not allow for the ferroelectric ground state with a spontaneous
polarization.

Numerically the problem can be easily solved since the numerical method
used for the spinless FKM can be straightforwardly generalized
also for the spin-one-half FKM. Unfortunately, including spins will
result in further reduction of the size of clusters that can be analyzed
using the exact-diagonalization method.
In order to compensate partially for the small size of clusters we next
examine the model only for strong $d$-$f$ interactions  that (as was shown
for the spinless FKM) minimize considerably the finite-size effects.

The Hamiltonian of the spin-one-half FKM can be obtained directly
from the spinless model by including the spins for both $d$ and $f$
electrons and by adding the on-site Coulomb interaction $U_{ff}$ that acts
between two $f$ electrons of opposite spins (the last term):

\begin{eqnarray}
H&=&\sum_{ij\sigma}t_{ij}d^+_{i\sigma}d_{j\sigma}+
U_{df}\sum_{i\sigma\sigma'}f^+_{i\sigma}f_{i\sigma}
d^+_{i\sigma'}d_{i\sigma'}+
E_f\sum_{i\sigma}f^+_{i\sigma}f_{i\sigma}+
V\sum_{i\sigma}d^+_{i\sigma}f_{i\sigma}+h.c. \nonumber \\
&&+\frac{U_{ff}}{2}\sum_{i\sigma}f^+_{i\sigma}f_{i\sigma}f^+_{i-\sigma}
f_{i-\sigma}.
\end{eqnarray}

The ground-state properties of this model, for $V=0$, have been investigated
in our preceding paper~\cite{F4}. We have found that numerical results
depend strongly on $f$-$f$ interaction strength $U_{ff}$, but they are
relatively insensitive to  $d$-$f$ interaction strength $U_{df}$ 
(for $U_{df} >2 $). 
Therefore to represent the typical behavior of the model at nonzero $V$, 
and to minimize finite-size effects we choose in the next study the value
$U_{df}=3$ that is sufficiently large to stabilize the system. Another
advantage of this selection is that the ground-state phase diagram of
the spin-one-half FKM without hybridization is well understood~\cite{F4}
for large values of $U_{df}$. Particularly,
in the strong-interaction limit $U_{ff}>4/\pi$  the ground state is
insulating for $E_f<-4/\pi$ and metallic for $E_f>-4/\pi$. At $E_f=-4/\pi$
the model exhibits a discontinuous insulator-metal transition
that is accompanied by a integer-valence transition from $n_f=1$ ($n_d=0$) to 
$n_f=0$ ($n_d=1$). The same behavior exhibits the model also in the opposite
limit $U_{ff}<4/\pi$: the ground state is insulating for $E_f<E_c(U_{ff})$
and metallic for $E_f>E_c(U_{ff})$. However, a discontinuous insulator-metal
transition that takes place at $E_f=E_c(U_{ff})$ realizes now between
an integer-valence state $n_f=1$ and an inhomogeneous intermediate-valence
state $n_f \neq 0$. These results show that there are only three physically
different ground states in the spin-one-half FKM without hybridization,
and namely, an insulating integer-valence ground state with
$n_f=1$, a metallic integer-valence ground state with $n_f=0$, and
a metallic intermediate-valence ground state with $0<n_f<1$.
Here we examine whether  these ground states are stable against
a small, finite hybridization or whether some new ground states are obtained
if one starts from a finite hybridization and studies the $V\to 0$
limit of the model. Again the special attention is devoted to the question
if the model can exhibit a ferroelectric ground state with a spontaneous,
nonvanishing polarization $P_{df}=<d^+_{i\sigma}f_{i\sigma}>$.

The strong-coupling ($U_{ff}=10$) numerical results for $P_{df}$ obtained
on finite clusters of 4 and 6 sites are displayed in Fig.~2. It is seen
that for both $E_f<-4/\pi$ (for $V=0$ an insulating integer-valence
state) and $E_f>-4/\pi$ (for $V=0$ a metallic integer-valence state)
the $<d^+f>$-expectation value vanishes in the limit $V\to 0$ indicating
that there is no spontaneous polarization in the spin-one-half FKM.
It should be noted that this result is expected since approximate
solutions~\cite{See} lead also to $P_{df}=0$ for the integer-valence
states with $n_f=1$ and $n_f=0$. A less trivial situation is expected
in the intermediate-valence state with $0<n_f<1$. As was discussed above
this state exists in the spin-one-half FKM without hybridization for
$U_{ff}<4/\pi$ and $E_f>E_c(U_{ff})$. Particularly, for $L=4,$ $U_{df}=3$ and
$U_{ff}=0.4$ we have found that $n_f=1$ for $E_f<-1.35$, $n_f=0.5$
for $-1.35<E_f<-0.65$, and $n_f=0$ for $E_f>-0.65$.
To examine the stability of intermediate-valence state  against a small,
finite hybridization we chose the value $E_f=-1$ for numerical calculations.
The results obtained for $P_{df}$ are shown in Fig.~3.
Again the $<d^+f>$-expectation value vanishes in the limit $V\to 0$
indicating the absence of spontaneous polarization. We have verified this
important result also for $L=8$ (see inset in Fig.~3).
Unfortunately, due to the memory limitations we were not able to continue
with numerical computations on larger lattices (the memory requirement of the
Lanczos method for clusters larger than $L=8$ is beyond the reach of
present day computers) to exclude definite the existence of spontaneous
polarization in the spin-one-half FKM. However, in accordance with results
obtained for the spinless FKM we do not expect that the behavior of $P_{df}$
for $V\to 0$ could be qualitatively changed on larger lattices.

In summary, we have used the extrapolation of small-cluster
exact-diagonalization calculations to study the possibility of electronic
ferroelectricity in the one dimensional spinless and spin-one-half FKM.
It was found that neither spinless nor spin-one-half version
of the FKM does not allow for a ferroelectric ground state with a
spontaneous polarization, i.e. there is no nonvanishing
$<d^{+}f>$-expectation value for vanishing hybridization $V$.

\vspace{0.5cm}
This work was supported by the Slovak Grant Agency for Science
under grant No. 4177/97.

\newpage

\centerline{\bf Figure Caption}

\vspace{0.5cm}

Fig.~1. Hybridization dependence of the $d$-$f$-polarization
$P_{df}=<d^+_if_i>$ in the spinless FKM calculated for three
different values of $U_{ff}$ and $L$. The symmetric case $E_f=0$.

\vspace{0.5cm}

Fig.~2. Hybridization dependence of the $d$-$f$-polarization
$P_{df}=<d^+_{i\sigma}f_{i\sigma}>$ in the spin-one-half FKM calculated
for $E_f=-2$ (for $V=0$ an insulating integer-valence
state) and $E_f=0$ (for $V=0$ a metallic integer-valence state).
Inset: Hybridization dependence of $P_{df}$ in the limit of vanishing $V$
for $L=4$ and $L=6$.

\vspace{0.5cm}

Fig.~3. Hybridization dependence of the $d$-$f$-polarization
$P_{df}=<d^+_{i\sigma}f_{i\sigma}>$ in the spin-one-half FKM calculated
for $E_f=-1$ (for $V=0$ an intermediate-valence state).
Inset: Hybridization dependence of $P_{df}$ in the limit of vanishing $V$
for $L=4$ and $L=8$.

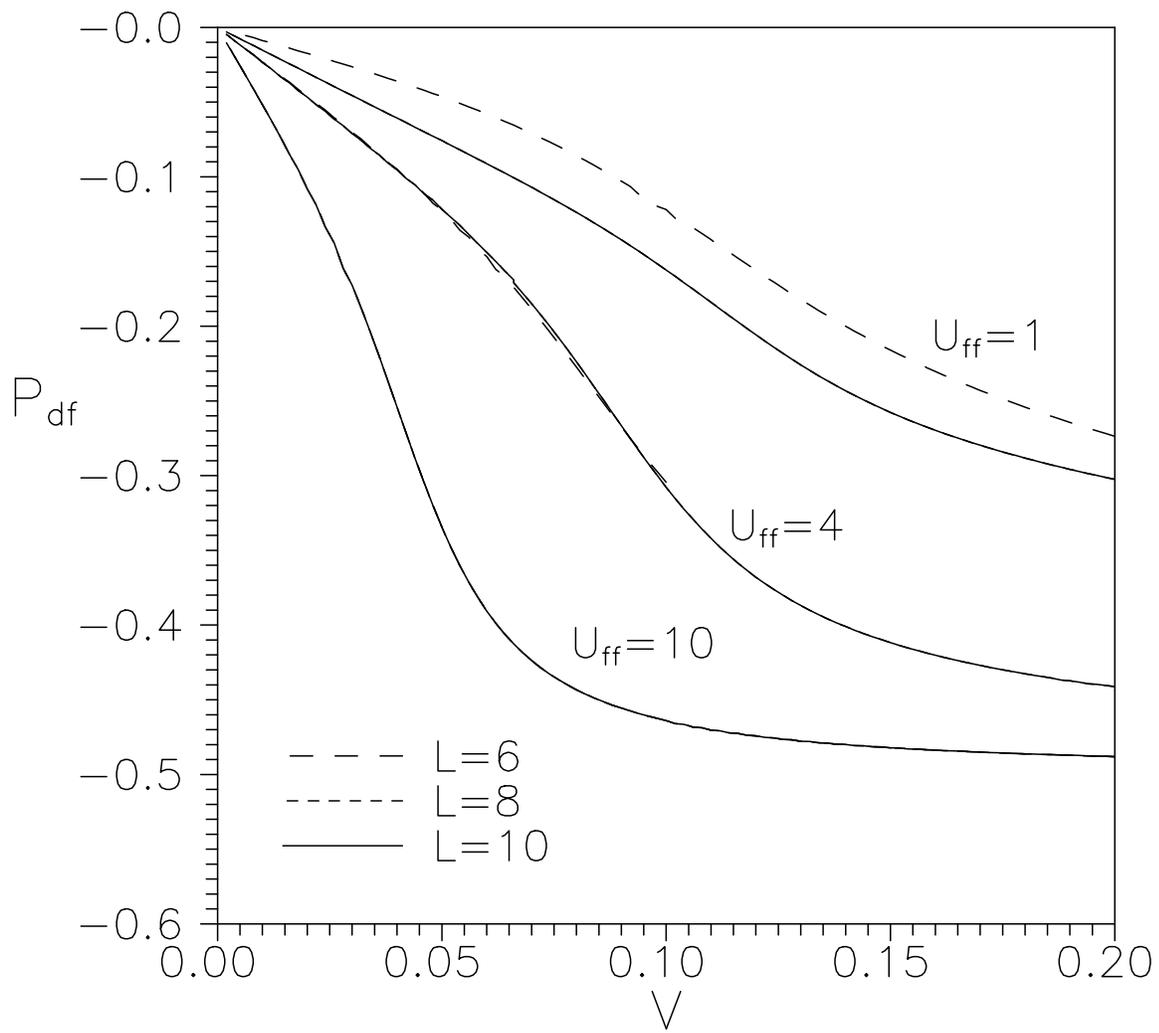
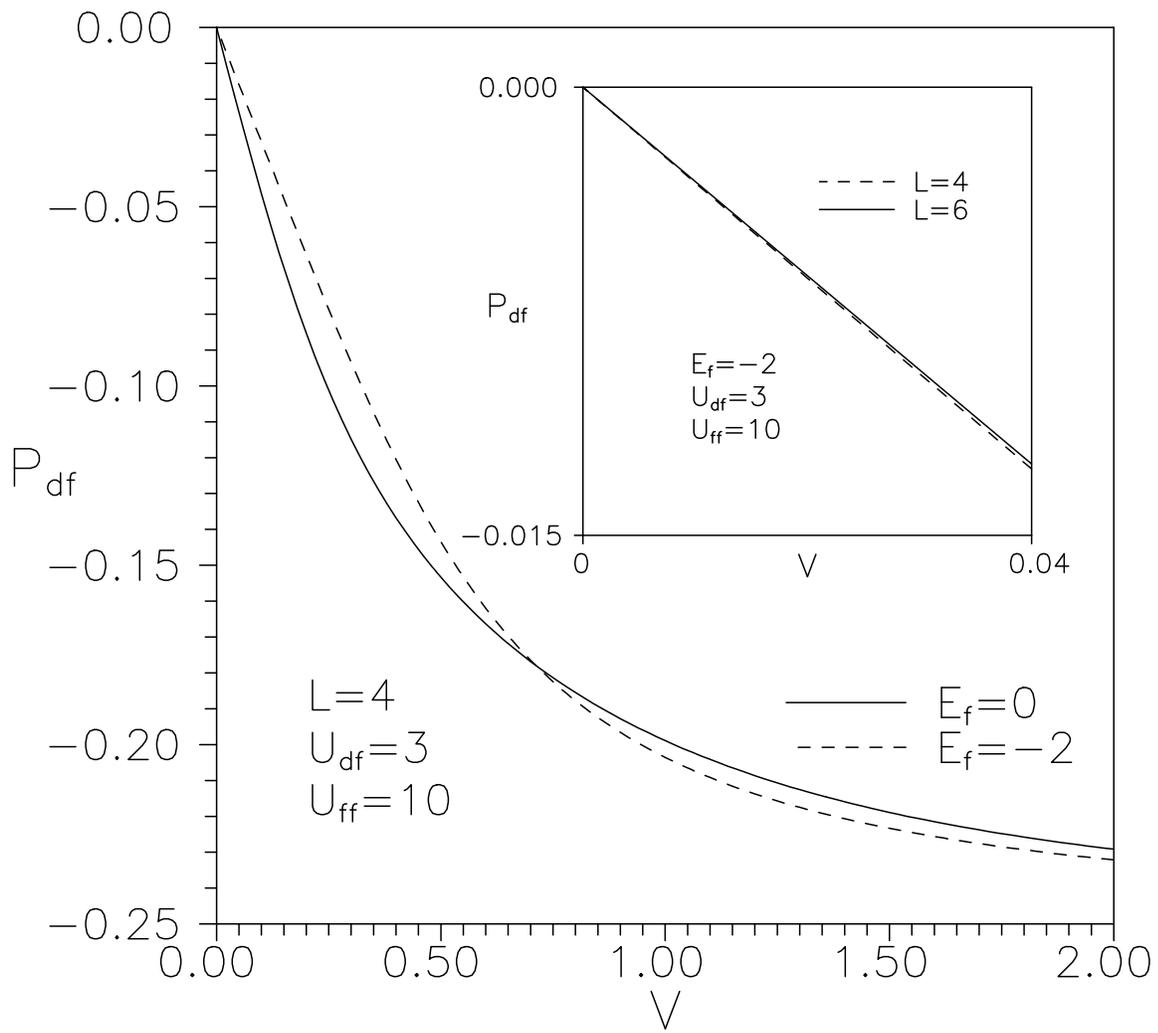
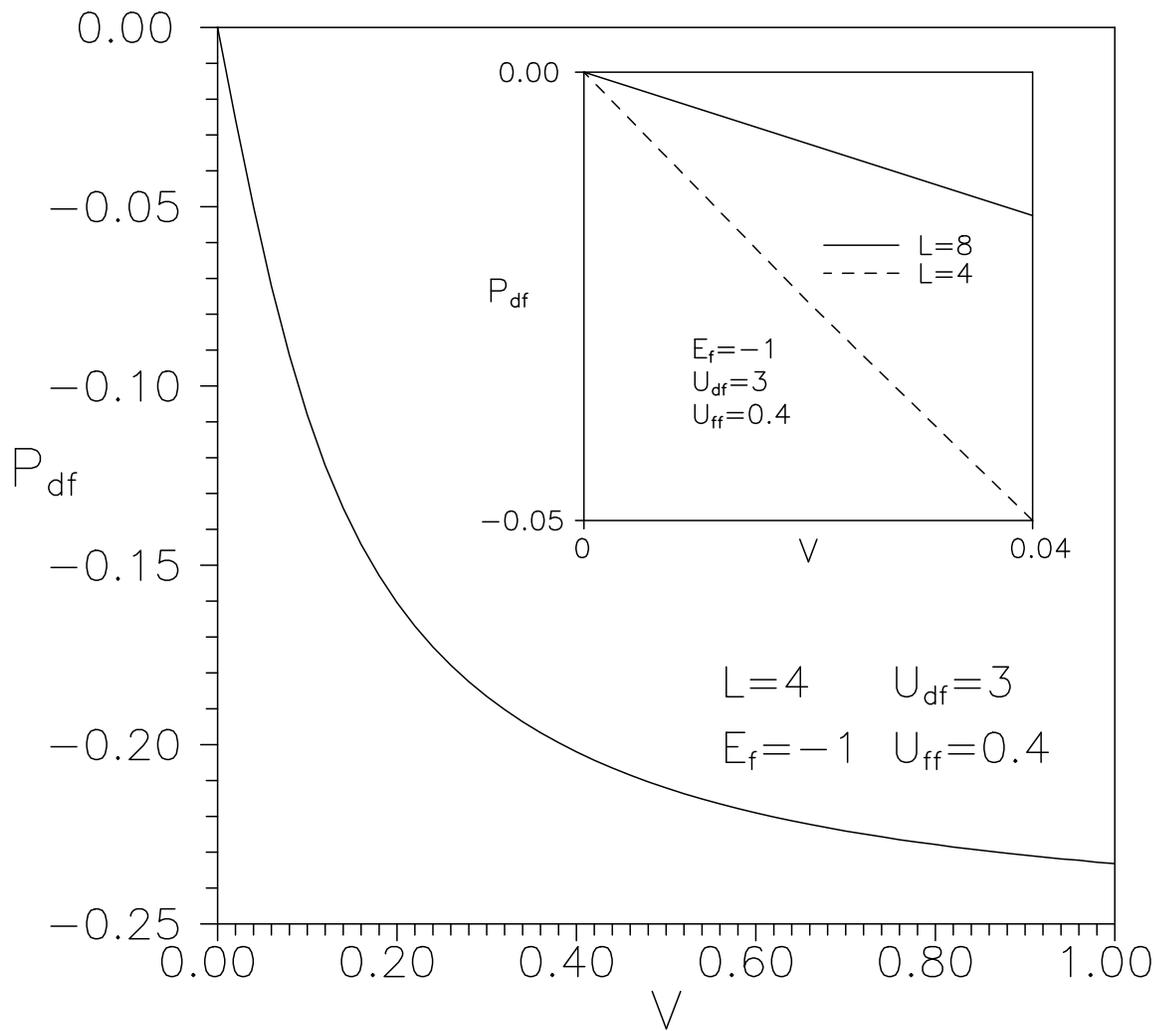
\end{document}